\documentclass[a4paper]{article}

\usepackage{INTERSPEECH2022}

\title{The SpeakIn Speaker Verification System for Far-Field Speaker Verification Challenge 2022}
\name{Yu Zheng\textsuperscript{*}, Jinghan Peng\textsuperscript{*}, Yihao Chen\textsuperscript{*}, Yajun Zhang, Jialong Wang, Min Liu, Minqiang Xu\textsuperscript{\dag}}
\address{SpeakIn Technologies Co. Ltd.}
\email{\{zhengyu, liumin, xuminqiang\}@speakin.ai}

\begin{document}

%
\maketitle
\begingroup\renewcommand\thefootnote{*}
\footnotetext{These authors share equal contribution to this work.}
\begingroup\renewcommand\thefootnote{\dag}
\footnotetext{Corresponding author.}
\begin{abstract}
This paper describes speaker verification (SV) systems submitted by the SpeakIn team to the Task 1 and Task 2 of the Far-Field Speaker Verification Challenge 2022 (FFSVC2022). SV tasks of the challenge focus on the problem of fully supervised far-field speaker verification (Task 1) and semi-supervised far-field speaker verification (Task 2). In Task 1, we used the VoxCeleb and FFSVC2020 datasets as train datasets. And for Task 2, we only used the VoxCeleb dataset as train set. The ResNet-based and RepVGG-based architectures were developed for this challenge. Global statistic pooling structure and MQMHA pooling structure were used to aggregate the frame-level features across time to obtain utterance-level representation. We adopted AM-Softmax and AAM-Softmax to classify the resulting embeddings. We innovatively propose a staged transfer learning method. In the pre-training stage we reserve the speaker weights, and there are no positive samples to train them in this stage. Then we fine-tune these weights with both positive and negative samples in the second stage. Compared with the traditional transfer learning strategy, this strategy can better improve the model performance. The Sub-Mean and AS-Norm backend methods were used to solve the problem of domain mismatch. In the fusion stage, three models were fused in Task1 and two models were fused in Task2. On the FFSVC2022 leaderboard, the EER of our submission is 3.0049\% and the corresponding minDCF is 0.2938 in Task1. In Task2, EER and minDCF are 6.2060\% and 0.5232 respectively. Our approach leads to excellent performance and ranks 1st in both challenge tasks.
  
\end{abstract}

\section{Introduction}


In the past several years, with the rapid development of the deep artificial neural network, near-field speaker verification (SV) performance has significantly improved\cite{snyder2017deep,vpr2014}. However, for the far-field speaker verification task, the quality of the speech signals is affected by energy decaying, reverberation, and environmental noise. All these factors make the far-field speaker verification problem more challenging. With intelligent speech assistants being used in smartphones and smart-home devices, more and more attention has been paid to the performance of far-field speaker recognition. The Far-Field Speaker Verification Challenge 2020 \cite{FFSVC2020} (FFSVC2020), which focused on both far-field text-dependent and text-independent speaker verification problems under noisy conditions in real scenarios, was launched to derive the development of far-field speaker recognition. FFSVC2020 successfully encouraged more researchers to pay attention to the far-field SV task. The Far-Field Speaker Verification Challenge 2022 (FFSVC2022) \cite{FFSVC2022_Eval_Plan}, heading in the same direction, still focuses on far-field speaker verification. The FFSVC2022 has two tasks, namely Task 1 and Task2. 
Task 1 is a fully supervised far-field speaker verification task that can only use the VoxCeleb dataset and FFSVC2020 dataset.
Task 2 focuses on cross-language self-supervised/semi-supervised learning, which is a novelty of this challenge.
In contrast to Task 1, the speaker label of the FFSVC2020 dataset cannot be used in Task 2. In addition to providing a supplementary set of the FFSVC2020 dataset, FFSVC2022 also provided a new far-field development and evaluation set collected in complex environments with multiple scenarios. 

The main difficulty of this challenge lies in cross-language and cross-domain. In response to this, we propose a new staged transfer training strategy that can greatly improve the performance of the systems on FFSVC2022 Task 1. We only use the close-talking English-based VoxCeleb data in the pre-training stage and preserve the speaker weights that need to be learned in the fine-tuning stage. In the fine-tuning stage, we add the far-field Mandarin-based FFSVC2020 data. The traditional transfer learning method does not reserve additional speaker weights in the pre-training stage, and randomly initializes these speaker weights in the fine-tuning stage.

All of our systems are deep-learning-based. Two different networks are used as encoders, namely ResNet and RepVGG. Based on the two-dimensional (2D) convolution layer, the networks get state-of-the-art performance for near-field speaker recognition in the reverberant and noisy environment. In this challenge, ResNet-based and RepVGG-based networks are used as encoders to generate frame-level representation. Each encoder is followed by a pooling layer to aggregate the frame-level features across time to obtain utterance-level representation. Global statistic pooling structure \cite{snyder2017deep} and multi-query multi-head attention (MQMHA) pooling structure \cite{zhao2021speakin} are used in our work. We use AM-Softmax and AAM-Softmax loss functions to increase the distance of inter-speakers and decrease the distance of the intra-speakers. We introduce the Sub-Center method \cite{deng2020sub} to reduce the influence of possible noisy samples. 
We use cosine similarity for scoring in both tasks. Adaptive score normalization (AS-Norm) \cite{asnorm} is used to increase the robustness against different channels and domains. For Task 2, we found that the Sub-Mean backend method drastically reduces the EER and minDCF. Finally, the scores from different single systems are weighted average fused. Our proposed system significantly outperforms the official baseline with 56.83\% and 27.33\% relative minDCF reductions on the Eval set of Task 1 and Task 2 respectively.

The paper is structured as follows: Section 2 describes the datasets used to train and test, feature extraction procedure, and augmentation methods. Section 3 introduces our network structure, including backbone, pooling structure, and loss function. Section 4 details the training methods of our systems. Section 5 focuses on the backend methods and fusion method. Experimental results are presented and analyzed in Section 6. Section 7 concludes this paper.

\section{Datasets}


\subsection{Training dataset}
For Task 1, VoxCeleb and FFSVC2020 datasets(train, dev, and supplementary set) were used to perform system development. VoxCeleb dataset contains 5994 speakers and the FFSVC2020 dataset has 155 speakers. We here adopted a 3-fold speed augmentation at first to generate extra twice speakers. Each speech segment in this dataset was perturbed by 0.9 and 1.1 factors based on the SoX speed function. Then we obtained total 18447 speakers, which is triple amount of the original speakers.

For Task 2, only VoxCeleb data with speed perturbation was used to perform system development. There are 17982 speakers in this dataset. FFSVC2020 dataset without speaker label information was used in the backend stage.

We applied the following techniques to augment each utterance:
\begin{itemize}
    \item  Reverberation: artificially reverberation using a convolution with simulated RIRs\cite{ko2017study} from the AIR dataset
    \item Music: taking a music file (without vocals) randomly selected from MUSAN\cite{snyder2015musan}, trimmed or repeated as necessary to match duration, and added to the original signal (5-15dB SNR).
    \item Noise: MUSAN noises were added at one-second intervals throughout the recording (0-15dB SNR).
    \item Babble: MUSAN speech was added to the original signal (13-20dB SNR).
\end{itemize}

We extracted 81-dimensional log Mel filter bank with energy based on Kaldi. The window size is 25 ms, and the frame shift is 10 ms. 200 frames of each feature were extracted without extra Voice Activation Detection (VAD). All features were cepstral mean normalized (CMN) in our training modes. 

\subsection{Development \& Evaluation dataset}
The development (Dev) data has the same data distribution as evaluation (Eval) data. All trial pairs of the development set and the evaluation set are single-channel speech segments. All enrollment utterances are close-talking speech segments recorded by telephone, while the test segments are close-talking or far-field audio recorded by tablet or telephone\cite{FFSVC2022_Eval_Plan}.



\section{Systems}


\subsection{ResNet}
As one of the most classical ConvNets, ResNet\cite{he2016deep} has proved its power in speaker verification. In our systems,  bottleneck-block-based ResNet (deeper structures:ResNet-74, ResNet-101, ResNet-152) are adopted. Base channels of all these ResNets are 64. We also implemented a deep and thin ResNet-221 structure which used ResNet\_v2\cite{DBLP:journals/corr/HeZR016} BottleNeck with 32 base channels.

\subsubsection{ResNet-D}
ResNet-D\cite{DBLP:journals/corr/abs-1812-01187} is a modification of the ResNet architecture that utilizes an average pooling tweak for downsampling. The motivation is that in the unmodified ResNet, the 1 × 1 convolution for the downsampling block ignores 3/4 of input feature maps. Such modification will not lead to the omission of information.

\subsubsection{fwSE}
A frequency-wise Squeeze-Excitation(fwSE) block\cite{https://doi.org/10.48550/arxiv.2109.04070}, which injects global frequency information across all feature maps, is used in our system.

\subsection{RepVGG}
In our previous work, we have proved that the RepVGG, as one of the re-parameterized models, shows competitive performance in speaker recognition\cite{zhao2021speakin, DBLP:journals/corr/abs-2110-09720}. We select RepVGG-A1, RepVGG-A2, and RepVGG-B1 as our backbones in this challenge. All RepVGG models adopt 64 base channels.

\subsection{Pooling Method}
The pooling layer aims to aggregate the variable sequence to an utterance level embedding. In addition to the global statistics pooling layer (GSP), we also used the multi-query multi-head attention pooling mechanism layer (MQMHA). RepVGG-A1 and RepVGG-A2 are followed by GSP, and other backbones are followed by MQMHA.

\subsection{Loss Function}
Recently, margin-based softmax methods have been widely used in speaker recognition works. To make a much better performance, we strengthen the AM-Softmax\cite{wang2018additive, wang2018cosface} and AAM-Softmax\cite{deng2019arcface} loss functions by the Sub-Center method.

The Sub-Center method \cite{deng2020sub} was introduced to reduce the influence of possible noisy samples. The formulation is given by:

\begin{equation}
  cos(\theta_{i,j})=\max_{1\leq k\leq K}(||x_i||\cdot||W_{j,k}||)
  \label{eq3}
\end{equation}

where the $\max$ function means that the nearest center is selected and it inhibits possible noisy samples interfering the dominant class center. $K$ means the number of sub-centers for each speaker class, and $k$ is the index of the sub-center.






\section{Training Protocol}
We used Pytorch\cite{paszke2019pytorch} to conduct our experiments. For Task 1, some of our models were trained through two stages, and the others were trained through an additional stage. And on Task 2, all of our systems were trained through the first stage.

\subsection{Stage 1: Pre-Training }

In the first stage, we used all VoxCeleb data with speed perturbation, consisting of 17982 speakers. It should be noted that we do not use any FFSVC2020 data at this stage. The number of classes is set to 17982 or 18447, depending on whether the weights of speakers from the FFSVC2020 dataset are reserved. The SGD optimizer with a momentum of 0.9 and weight decay of 1e-3 was used. We used 8 GPUs with 128 mini-batch and an initial learning rate of 0.08 to train all of our models. 200 frames of each sample in one batch were adopted. We adopted the ReduceLROnPlateau scheduler with a frequency of validating every 2,000 iterations, and the patience is 2. The minimum learning rate is 1.0e-6, and the decay factor is 0.1. All the models were trained with AM-Softmax in the first stage. Furthermore, the margin gradually increases from 0 to 0.2 \cite{liu2019large}. 

We propose a novel training method, which preserves the speaker weights in the pre-training stage that need to be learned for the fine-tuning stage, even though there are no positive samples to train these preserved weights in the pre-training phase. 
For example, the number of speakers in VoxCeleb dataset with speed perturbation is 17982, and we set the number of classes in loss function to 18447. In the pre-training stage, there are no positive samples but only negative samples to train these preserved weights of 465 classes. The preserved 465 classes correspond to 465 speakers in the FFSVC2020 dataset with speed perturbation, respectively.
When the number of classes is set to 17982, the weights of speakers of the FFSVC2020 dataset will be generated by random initialization in the second-stage training. However, preserving the speaker weights of the FFSVC2020 dataset in stage 1, we could use the trained weights to initialize in the fine-tuning stage. The experimental results confirm that this training method greatly improves the performance of the model, compared with the method of randomly initializing the weights of the last layer in traditional transfer learning. For specific experimental results, we can refer to the results of whether fwSE-ResNet34-D reserves weights in Table 1. 

\subsection{Stage 2: Fine-Tuning }

We used VoxCeleb dataset without speed perturbation and the FFSVC2020 dataset with speed perturbation to fine-tune all systems in the second stage. The training dataset consists of 6459 speakers. All the configurations and hyper-parameters were the same as in the first stage except the initial learning rate, which was 2e-5. We removed the speed augmented part from the VoxCeleb dataset, and discarded the corresponding weights at the same time. For this reason, the number of classes is changed to 6459.

\subsection{Stage 3: Large-Margin Fine-Tuning }

Large-Margin Fine-Tuning (LM-FT) \cite{thienpondt2020idlab} helps to further improve model performance for some of our models. We chose the second-stage model to fine-tune for an additional epoch. In the LM-FT stage, settings are slightly different from the second stage. Firstly, we only used the FFSVC2020 dataset set as the training data, removing the speed augmented part from the training set to avoid domain mismatch. Secondly, we changed the chunk size from 200 to 400 and increased the margin exponentially from 0.2 to 0.5. The AM-Softmax loss was replaced by AAM-Softmax loss. We found that the large-margin-based fine-tuning in the third stage is not stable. For some large models, extra large-margin-based fine-tuning after the second stage may make the model performance worse. As a result, we only do the third stage fine-tuning on some models, such as RepVGG-A1 and RepVGG-A2.

\section{Backend}
We used cosine distance for scoring in both Task 1 and Task 2. In addition, adaptive symmetric score normalization (AS-norm) \cite{asnorm} was used for Task 1, and Sub-Mean was used for Task 2.

\subsection{AS-Norm}
For Task 1, AS-Norm was used for all of the models. For AS-Norm, we selected the original VoxCeleb and FFSVC2020 dataset without any augmentation. The cohort was created by using the speaker's random one utterance embedding vector as a speaker center and consisted of 6149 speaker centers. Only part of the cohorts are selected to compute mean and standard deviation for normalization, and top-300 highest scores are selected for Task 1. 

\subsection{Sub-Mean }
Sub-Mean was used for models trained for Task 2. We randomly chose 40000 utterances from the FFSVC2022 dataset, then extracted the embedding vectors to compute the global mean embedding vector. The enrollment and test embedding vectors both subtract the mean embedding vector before scoring:
\begin{equation}
s(\bm{x_{e}} ,\bm{x_{t}}) = cos(\bm{x_{e}}-\bm{\bar{x}} ,\bm{x_{t}}-\bm{\bar{x}})
\end{equation}

where $\bm{x_{e}}$, $\bm{x_{t}}$ are enrollment and test speaker embedding vectors respectively, and $\bm{\bar{x}}$ is the mean embedding vector of 40000 utterances randomly chosen from the FFSVC2022 dataset.

\subsection{Fusion }
The results of all systems were fused using Logistic Regression on FFSVC2022 Dev set. We got the weight of each system and then selected the dominant systems to assign weights artificially. In the end, fusion was performed by computing the weighted average of the scores of selected individual systems.

\section{Results}


\begin{table*}[t]
\centering
\caption{Performance on FFSVC2022 Dev\&Eval set in Task 1. All systems used the AS-Norm backend method. Reserved Weight is T means that the corresponding model reserved the speaker weights of the FFSVC2020 dataset in the first stage, and F means that the corresponding model does not. S7 and S9 systems were trained in the third stage, the others were only trained in the second stage.}
\label{tab:tabel_two}
\setlength{\tabcolsep}{7.5mm}{
\resizebox{\textwidth}{!}{
\begin{tabular}{ccccccc}
\hline
System Index & System & Reserved Weight & \multicolumn{2}{c}{Dev} & \multicolumn{2}{c}{Eval} \\ \cline{4-7} 
 &  &  & EER(\%) & minDCF & EER(\%) & minDCF \\ \hline
S1 & fwSE-ResNet34-D & F & 3.4278 & 0.3830 & - & - \\
S2 & fwSE-ResNet34-D & T & 5.8806 & 0.5379 & - & - \\
S3 & ResNet74 & T & 3.3361 & 0.3621 & - & - \\
S4 & fwSE-ResNet101-D & T & 3.0694 & 0.3319 & - & - \\
S5 & ResNet152 & T & 2.6667 & 0.2939 & 3.1897 & 0.3108 \\
S6 & ResNet221\_v2(thin) & T & 2.9861 & 0.3237 & 3.3333 & 0.3307 \\
S7 & RepVGG-A1 & T & 3.8417 & 0.3910 & 4.1109 & 0.3933 \\
S8 & RepVGG-A2 & T & 3.4472 & 0.3554 & 3.7269 & 0.3699 \\
S9 & RepVGG-A2 & T & 3.5389 & 0.3536 & 3.7178 & 0.3651 \\
S10 & RepVGG-B1 & T & 3.2861 & 0.3247 & - & - \\ \hline
Fusion &  &  &  &  &  &  \\ \hline
S5,S6,S9 &  &  & \textbf{2.5000} & \textbf{0.2735} & \textbf{3.0049} & \textbf{0.2938} \\ \hline
\end{tabular}
}
}
\end{table*}

\begin{table*}[t]
\centering
\caption{Performance on FFSVC2022 Dev\&Eval set in Task 2. All models were trained only using the VoxCeleb dataset in stage 1. S2-S8 systems used the Sub-Mean backend method while the S1 system did not.}
\label{tab:tabel_two}
\setlength{\tabcolsep}{7.5mm}{
\resizebox{\textwidth}{!}{

\begin{tabular}{ccccccc}
\hline
System Index & System & Sub-Mean & \multicolumn{2}{c}{Dev} & \multicolumn{2}{c}{Eval} \\ \cline{4-7} 
 &  &  & EER(\%) & minDCF & EER(\%) & minDCF \\ \hline
S1 & fwSE-ResNet34-D & F & 8.6861 & 0.6901 & - & - \\
S2 & fwSE-ResNet34-D & T & 7.2444 & 0.5650 & - & - \\
S3 & fwSE-ResNet101-D & T & 7.1472 & 0.5617 & - & - \\
S4 & ResNet152 & T & 6.4639 & 0.5132 & 6.6924 & 0.5374 \\
S5 & ResNet221\_v2(thin) & T & 6.4417 & 0.5399 & - & - \\
S6 & RepVGG-A1 & T & 7.4306 & 0.5929 & - & - \\
S7 & RepVGG-A2 & T & 6.9583 & 0.5600 & - & - \\
S8 & RepVGG-B1 & T & 7.0861 & 0.5556 & - & - \\ \hline
Fusion &  &  &  &  &  &  \\ \hline
S4,S5 &  &  & \textbf{5.9833} & \textbf{0.5004} & \textbf{6.2060} & \textbf{0.5232} \\ \hline
\end{tabular}

}
}
\end{table*}

Results of experiments on all our systems developed for the Task 1 and Task 2 are displayed in Table 1 and Table 2 respectively. The performance is measured on the FFSVC2022 development and evaluation set in terms of Equal Error Rate (EER) and Minimum Detection Cost(minDCF) with a prior target probability, $P_{tar}$ of 0.01. All systems in Table 1 are the results of models trained in stage 2 or stage 3 for Task 1, while all systems in Table 2 are the results of models trained in stage 1 only using the VoxCeleb dataset for Task 2.

All systems in Table 1 are the results of the cosine score after AS-Norm calibration for Task 1. Among all the single systems, ResNet152 gets the best performance by both EER and minDCF, which has a 6.6924\% EER and 0.5374 minDCF after AS-Norm. The final fusion result used three single systems and get a 3.0049\% EER and 0.2938 minDCF. 
The S1 system that reserves the speaker classification weights of the FFSVC2020 dataset in the first training stage gets a much lower minDCF value than the S2 system that does not. According to the results of the S1 system and S2 system, the reserving weight training strategy results in a relative minDCF reduction of 28.8\%. The other systems (S3-S10) all reserve weights in the first stage. 
The S8 system was trained in the second stage, and the S9 system was trained in the third stage. Comparing the S8 system and S9 system, we conclude that an additional Large-Margin Fine-Tuning can lower the minDCF slightly. However, we found that for large models (such as ResNet152 and RepVGG-B1) an extra LM-FT probably degraded model performance. Therefore we did not perform LM-FT on all individual systems.

For Task 2, all the models were trained in stage 1. S2-S8 systems used the Sub-Mean method described in Section 5.2 while the S1 system did not. Comparing the S1 system and S2 system, the Sub-Mean method brings 18.13\% relative minDCF reductions. Same as Task 1, the best individual system is ResNet152, which has a 6.6924\% EER and 0.5374 minDCF. The EER and minDCF of the fused result are 6.2060\% and 0.5232 respectively. In our experiments, we found that AS-Norm calibration after Sub-Mean does not further reduce the minDCF value. As a result, we only use the Sub-Mean backend method for Task 2. Finally, our system won the first place in both challenge tasks.


\section{Conclusions}
We experimented with multiple models on the SV, and ResNet achieved the best results on both tasks. Data augmentation, hyper-parameter changes in the fine-tuning stage, and score normalization in the backend have all brought improvement. Experimental results show that the larger model outperforms the small model on both tasks of the FFSVC2022.

Training in stages with using different train dataset and strategies in different stages can greatly improve model performance. The model trained using only VoxCeleb data will perform slightly worse than the model trained using VoxCeleb and FFSVC2020 data in the first stage. But the former will perform far better than the latter after fine-tuning in the second stage. 

From the experimental results, we conclude that reserving the weights of the speakers from the FFSVC2020 dataset with only using the VoxCeleb dataset to train in the first stage makes the model perform much better after fine-tuning in the second stage. We speculated that the randomly initialized weights of speakers of the FFSVC2020 dataset are hard to converge possibly due to the small learning rate in the second stage. 
We found that an additional large-margin-based fine-tuning after the second-stage fine-tuning can further improve the model performance, such as RepVGG-A1 and RepVGG-A2. However, our experimental results show that this method is unstable. For some large models, such as RepVGG-B1 and ResNet152, the refinement may make the model performance worse.

For Task 2, the speaker labels of the far-field FFSVC2020 dataset cannot be used in training. We only use the close-talking VoxCeleb dataset to train our models. Instead, the development set and evaluation set are far-field audio datasets. Therefore, there is a domain mismatch between training data and evaluation data. We focused on the solution of domain mismatch and used the Sub-Mean to solve the problem. Our experimental results show that Sub-Mean is pretty useful for domain mismatch.

\bibliographystyle{IEEEtran}
\bibliography{INTERSPEECH2022_BibEntries}

\end{document}